\documentclass[a4paper]{article}
\usepackage{epsfig,amsmath,amsfonts,amssymb,setspace,multirow,textcomp,xspace,
url}
\usepackage[T1]{fontenc}
\usepackage[sort&compress,numbers]{natbib}

\makeatletter
\renewcommand{\@oddhead}{\textit{Advances in Astronomy and Space Physics} \hfil}
\renewcommand{\@evenfoot}{\hfil \thepage \hfil}
\renewcommand{\@oddfoot}{\hfil \thepage \hfil}
\makeatother

\renewenvironment{thebibliography}[1]{\begin{oldthebibliography}{#1}\setlength{
\parskip}{0ex}\setlength{\itemsep}{0ex}}{\end{oldthebibliography}}
\newcommand{\xmm}{\textit{XMM-Newton}\xspace}
\newcommand{\chan}{\textit{Chandra}\xspace}
\newcommand{\dm}{{\textsc{dm}}} 
\renewcommand{\S}{\mathcal{S}} 

\usepackage{tabularx}

\usepackage{booktabs} 
\newcommand {\otoprule }{\midrule [\heavyrulewidth]} 
\newcommand{\suza}{\textsl{Suzaku}\xspace} 

\begin{document}
\fontsize{11}{11}\selectfont %

\title{New emission line at $\sim$3.5~keV -- observational status, connection with radiatively decaying 
dark matter and directions for future studies}
\author{\textsl{D.\,A.~Iakubovskyi$^{1}$}}
\date{\vspace*{-6ex}}
\maketitle
\begin{center} {\small $^{1}$Bogolyubov Institute of Theoretical Physics,
Metrologichna str. 14-b, 03680, Kyiv, Ukraine\\
{\tt yakubovskiy@bitp.kiev.ua}}
\end{center}

\begin{abstract}
Recent works of~\cite{Bulbul:14a,Boyarsky:14a}, claiming the detection 
of the extra emission line with energy $\sim$3.5~keV in 
X-ray spectra of certain clusters of galaxies and nearby Andromeda galaxy,
have raised considerable interest in astrophysics 
and particle physics communities. A number of new observational studies
claim detection or non-detection of the extra line in X-ray spectra
of various cosmic objects. 
In this review I summarize existing results of these studies, overview possible interpretations
of the extra line, including intriguing connection with radiatively decaying 
dark matter, and show future directions achievable with existing and planned X-ray cosmic missions.
\\[1ex]
\end{abstract}

%

\section*{\sc Introduction}

\indent \indent We still have to explore the origin of \emph{dark matter} -- gravitationally interacting 
substance which constitutes the major fraction of non-relativistic matter in the Universe.
None of known elementary particles constitutes the bulk of dark matter. Despite that fact, the most 
plausible\footnote{Viable alternatives include modified laws of gravity and/or Newtonian dynamics 
(see e.g.~\cite{Bekenstein:10,Moffat:11,Famaey:11,Milgrom:14}), primordial black holes 
(see e.g.~\cite{Carr:05,Capela:12}) etc.}
hypothesis is the dark matter made of elementary particles which implies the extension of
Standard Model of particle physics and is of considerable interest for particle physicists.

Dozens of Standard Model extensions proposed so far range by main parameters -- mass of dark matter particles 
and their interaction strength with Standard Model particles -- by tens of orders of magnitude.
Astrophysical observations of dark matter objects can probe some of them.
An interesting example is \emph{radiatively decaying dark matter}. 
If a dark matter particle interacts with
electrically charged Standard Model particles, it usually decays\footnote{A widely-known 
examples where this is \emph{not} the case are dark matter particles holding a new quantum number 
\emph{conserved} by Standard Model interactions, such as R-parity for supersymmetric models, Kalutza-Klein 
number for extra dimensions, etc. In this case, dark matter decays are exactly forbidden by special structure 
of the theory, and the main astrophysical effect for such dark matter candidates is \emph{annihilation}
of dark matter particles with their antiparticles.} emitting a photon. 
For 2-body radiative channel, Doppler broadening of dark matter in haloes will cause a 
narrow \emph{dark matter decay line}.
Such a line possesses a number of specific properties allowing to
distinct it from astrophysical emission lines or instrumental line-like features:
\begin{itemize}
 \item its position in energy is solely determined by the mass of dark matter particle and the redshift of dark 
 matter halo (i.e. if one neglects the mass of other decay product, the line position is 
 $\frac{m_\dm c^2}{2(1+z)}$), having different scaling with redshift $z$ compared with instrumental features; 
 \item its intensity should be proportional to \emph{dark matter column density}
 \footnote{Dark matter column densities for different dark matter-domianted objects are 
 compiled in~\cite{Boyarsky:09b}.} $\S_\dm=\int \rho_\dm d\ell$; 
 due to different 3D distributions of dark and visible matter, comparison
 of line intensity within a given object and among different objects allows to choose between decaying 
 dark matter and astrophysical origin of the line; 
 \item it is broadened with characteristic velocity of dark matter usually 
 different from that of visible matter.
\end{itemize}
The above properties allow to reliably establish that the line comes from decaying dark matter. 
In other words, we can \emph{directly detect} radiatively decaying dark matter
relying on astrophysical measurements.

The search for decaying dark matter in X-rays lasts for about a decade starting from pioneering proposals 
in~\cite{Abazajian:01b,Dolgov:00}. The searches prior to February~2014 
are summarized in Table~I of~\cite{Neronov:13}; the only exception is recent study of~\cite{Horiuchi:13}. 
These searches have not revealed the presence
of \emph{viable} candidate lines from decaying dark matter and obtained only upper bounds on radiative decay 
lifetime of dark matter particles. The only exception is the claim of~\cite{Prokhorov:10}
about the excess of Fe~XXVI  Lyman-$\gamma$ line at $8.7$~keV in \suza\ spectrum of Galactic 
Center~\cite{Koyama:06} compared with the standard ionization and recombination processes.
Existing X-ray telescopes do not allow to reach any reliable conclusion about 
the nature of this excess, so the claim of~\cite{Prokhorov:10} should be tested with new instruments
having better spectral resolution, discussed in e.g.~\cite{Boyarsky:12c}.

\begin{table}
    \begin{tabularx}{\textwidth}{lXccccc}
      \toprule
 Ref. & Object & Redshift & Instrument & Exposure, & Line position, & Line flux,  \\
      &        &        &              & Msec      & keV            & $10^{-6}$ ph/sec/cm$^{2}$ \\
      \otoprule
 \cite{Bulbul:14a} & Full stacked sample & 0.009-0.354 & MOS & 6 & 3.57$\pm$0.02 & 4.0$\pm$0.8 \\
 \cite{Bulbul:14a} & Full stacked sample & 0.009-0.354 & PN & 2 & 3.51$\pm$0.03 & 3.9$^{+0.6}_{-1.0}$ \\
 \cite{Bulbul:14a} & Coma+Centaurus+Ophiuchus & 0.009-0.028 & MOS & 0.5 & 3.57$^a$ & 15.9$^{+3.4}_{-3.8}$ \\
 \cite{Bulbul:14a} & Coma+Centaurus+Ophiuchus & 0.009-0.028 & PN & 0.2 & 3.57$^a$ & $< 9.5$ (90\%) \\
 \cite{Bulbul:14a} & Perseus ($<$ 12') & 0.016 & MOS & 0.3 & 3.57$^a$ & 52.0$^{+24.1}_{-15.2}$ \\
 \cite{Bulbul:14a} & Perseus ($<$ 12') & 0.016 & PN & 0.05 & 3.57$^a$ & $< 17.7$ (90\%) \\
 \cite{Bulbul:14a} & Perseus (1-12') & 0.016 & MOS & 0.3 & 3.57$^a$ & 21.4$^{+7.0}_{-6.3}$ \\
 \cite{Bulbul:14a} & Perseus (1-12') & 0.016 & PN & 0.05 & 3.57$^a$ & $< 16.1$ (90\%) \\ 
 \cite{Bulbul:14a} & Rest of the clusters & 0.012-0.354 & MOS & 4.9 & 3.57$^a$ & 2.1$^{+0.4}_{-0.5}$ \\
 \cite{Bulbul:14a} & Rest of the clusters & 0.012-0.354 & PN & 1.8 & 3.57$^a$ & 2.0$^{+0.3}_{-0.5}$ \\ 
 \cite{Bulbul:14a} & Perseus ($>$ 1') & 0.016 & ACIS-S & 0.9 & 3.56$\pm$0.02 & 10.2$^{+3.7}_{-3.5}$ \\
 \cite{Bulbul:14a} & Perseus ($<$ 9') & 0.016 & ACIS-I & 0.5 & 3.56$^a$ & 18.6$^{+7.8}_{-8.0}$ \\ 
 \cite{Bulbul:14a} & Virgo ($<$ 500'') & 0.003-0.004 & ACIS-I & 0.5 & 3.56$^a$ & $< 9.1$ (90\%) \\
\midrule
 \cite{Boyarsky:14a} & M31 ($<$ 14') & -0.001$^b$ & MOS & 0.5 & 3.53$\pm$0.03 & 4.9$^{+1.6}_{-1.3}$ \\
 \cite{Boyarsky:14a} & M31 (10-80') & -0.001$^b$ & MOS & 0.7 & 3.50-3.56 & $< 1.8$ ($2\sigma$) \\
 \cite{Boyarsky:14a} & Perseus (23-102') & 0.0179$^b$ & MOS & 0.3 & 3.50$\pm$0.04 & 7.0$\pm$2.6 \\
 \cite{Boyarsky:14a} & Perseus (23-102') & 0.0179$^b$ & PN & 0.2 & 3.46$\pm$0.04 & 9.2$\pm$3.1 \\
 \cite{Boyarsky:14a} & Perseus, 1st bin (23-37') & 0.0179$^b$ & MOS & 0.2 & 3.50$^a$ & 13.8$\pm$3.3 \\
 \cite{Boyarsky:14a} & Perseus, 2nd bin (42-54') & 0.0179$^b$ & MOS & 0.1 & 3.50$^a$ & 8.3$\pm$3.4 \\
 \cite{Boyarsky:14a} & Perseus, 3rd bin (68-102') & 0.0179$^b$ & MOS & 0.03 & 3.50$^a$ & 4.6$\pm$4.6 \\
 \cite{Boyarsky:14a} & Blank-sky & --- & MOS & 7.8 & 3.45-3.58 & $< 0.7$ ($2\sigma$) \\ 
\midrule
 \cite{Riemer-Sorensen:14} & Galactic center (2.5-12') & 0.0 & ACIS-I & 0.8 & $\simeq 3.5$ & $\lesssim 25$ ($2\sigma$) \\ 
\midrule
 \cite{Jeltema:14} & Galactic center (0.3-15') & 0.0 & MOS & 0.7 & $\simeq 3.5$ & $< 41$ \\ 
 \cite{Jeltema:14} & Galactic center (0.3-15') & 0.0 & PN & 0.5 & $\simeq 3.5$ & $< 32$ \\ 
 \cite{Jeltema:14} & M31 & 0.0 & MOS & 0.5 & 3.53$\pm$0.07 & 2.1$\pm$1.5$^c$ \\ 
\midrule
\cite{Boyarsky:14b} & Galactic center ($<$ 14') & 0.0 & MOS & 0.7 & 3.539$\pm 0.011$ & 29$\pm$5 \\ 
\midrule
\cite{Malyshev:14} & Combined dSphs & 0.0 & MOS+PN & 0.4+0.2 & 3.55$^a$ & $< 0.254$ (90\%) \\ 
\midrule
 \cite{Anderson:14} & Combined galaxies ($\gtrsim 0.01 R_{vir}$) & 0.0 & MOS & 14.6 & $\simeq 3.5$ & unknown$^d$ \\ 
 \cite{Anderson:14} & Combined galaxies ($\gtrsim 0.01 R_{vir}$) & 0.0 & ACIS-I & 15.0 & $\simeq 3.5$ & unknown$^d$ \\ 
\midrule
\cite{Urban:14} & Perseus core ($<$ 6') & 0.0179$^b$ & XIS & 0.74 & 3.510$^{+0.023}_{-0.008}$ & $32.5^{+3.7}_{-4.3}$ \\ 
\cite{Urban:14} & Perseus confined (6-12.7') & 0.0179$^b$ & XIS & 0.74 & 3.510$^{+0.023}_{-0.008}$ & $32.5^{+3.7}_{-4.3}$ \\ 
\cite{Urban:14} & Coma ($<$ 12.7') & 0.0231$^b$ & XIS & 0.164 & $\simeq 3.45^e$ & $\simeq 30^e$ \\ 
\cite{Urban:14} & Ophiuchus ($<$ 12.7') & 0.0280$^b$ & XIS & 0.083 & $\simeq 3.45^e$ & $\simeq 40^e$ \\ 
\cite{Urban:14} & Virgo ($<$ 12.7') & 0.0036$^b$ & XIS & 0.09 & 3.55$^a$ & $< 6.5$ (2$\sigma$) \\ 

\bottomrule
 \end{tabularx}
\caption{Properties of $\sim$3.5~keV line searched in different X-ray datasets observed by MOS and PN cameras
on-board \xmm\ observatory, ACIS instrument on-board \chan\ observatory and XIS instrument on-board \suza\ 
observatory. All error bars are at 1$\sigma$
 (68\%) level.
   \newline
   $^a$ Line position was fixed at given value.
   \newline
   $^b$ Redshift was fixed at NASA Extragalactic Database (NED) value.
   \newline
   $^c$ The line was detected at $< 90$\% confidence level. Such a low flux (compared with~\cite{Boyarsky:14a}) 
   was due to unphysically enhanced level of continuum at 
   3-4~keV band used in~\cite{Jeltema:14}, see~\cite{Boyarsky:14c} for details.
   \newline
   $^d$ \cite{Anderson:14} only qouted the ``minimal probed values'' of sterile neutrino 
   mixing angle $\sin^2(2\theta) \lesssim 2\times 10^{-11}$ by \xmm/MOS and 
   $\lesssim 5\times 10^{-11}$ by \chan/ACIS-I. For \xmm\ dataset with average dark matter column density
   in field-of-view equal to 100~$M_\odot$/pc$^2$, these values would correspond to upper bound 
   on $\sim$3.5~keV line 
   flux $\sim 3.0\times 10^{-7}$ and $\sim 7.5\times 10^{-7}$ ph/sec/cm$^2$, respectively.
   \newline
   $^e$ Parameters estimated from Fig.~3 of~\cite{Urban:14}, see text. 
   }\label{tab:line-properties}
\end{table}


\section*{\sc Observational status of $\sim$3.5~keV line}

In February 2014, the situation has changed dramatically: two groups~\cite{Bulbul:14a,Boyarsky:14a}
have claimed the presence of an extra line at $\sim$3.5~keV. Stacking X-ray spectra of central parts of 81 galaxy 
clusters (observed by \xmm\ and \chan) in emitter's rest frame allowed~\cite{Bulbul:14a} to reach unprecedented 
sensitivity, compared to previous line searches in galaxy clusters. 
As a result, new line has been detected in independent subsets -- Perseus galaxy cluster,
the sum of three nearby galaxy clusters (Coma, Centaurus and Ophiuchus), and the rest of galaxy clusters of 
their sample. 
On the other hand, \cite{Boyarsky:14a} presented analysis of several \emph{independent} datasets -- 
the nearby Andromeda galaxy, outskirts of the Perseus cluster and the new blank-sky dataset -- observed by \xmm, 
and claimed the detection of new line in Perseus outskirts (using set of observations completely different from 
that of~\cite{Bulbul:14a}). After that, the same group~\cite{Boyarsky:14b} has presented another evidence 
for extra line at $\sim$3.5~keV by looking at the central part of our Galaxy.
Other recent study~\cite{Urban:14} detected the 3.5~keV line in the central part of the Perseus 
cluster\footnote{\cite{Urban:14} also analyzed \suza\ observations of Coma, Ophiuchus and Virgo galaxy clusters.
In fact, the faint extra line at $\sim$3.45~keV (rest-frame) was found in Coma and Ophiuchus spectra, see their 
Fig.~3. Because the position of this extra line line coincides with other detections within \suza\ energy 
resolution ($\simeq 150$~eV), we included the detections in Coma and Ophiuchus into 
Table~\ref{tab:line-properties}. Taking into account this fact and the level actual uncertainty between X-ray 
and weak lensing modeling at virial radius~\cite{Okabe:14}, the results of~\cite{Urban:14} are consistent with 
decaying dark matter hypothesis.} observed by \suza.
These studies have been accompanied by claims
of several other groups~\cite{Riemer-Sorensen:14,Jeltema:14,Malyshev:14,Anderson:14} that \emph{have not}
detected the extra line at $\sim$3.5~keV in several different datasets of dark matter objects.
Basic properties of all these datasets are summarized in Table~\ref{tab:line-properties}.

\section*{\sc Possible explanations}

\indent \indent The following possibilities 
for the origin of new line have been considered:
%

1. The possibility that new line is \emph{not} from astrophysical emission has first been studied 
in pioneering papers~\cite{Bulbul:14a,Boyarsky:14a}. Using detailed computations of line intensities in thermal 
plasma hosted by galaxy clusters based on atomic line database ATOMDB v.2.0.2~\cite{AtomDB}, \cite{Bulbul:14a} 
argued that possible contributions from astrophysical lines near 3.5~keV are factor $\gtrsim 30$ 
smaller than the detected flux of the extra line. In addition, \cite{Boyarsky:14a} showed that
the angular distribution of $\sim 3.5$~keV line in Perseus galaxy outskirts is much better consistent
with decaying dark matter distribution than that with astrophysical emission.

But, these conclusions of~\cite{Bulbul:14a,Boyarsky:14a} were questioned in~\cite{Jeltema:14}, in which 
the authors argue that a) it is possible to explain new line in central part of our Galaxy  and in combined 
dataset of~\cite{Bulbul:14a} with contibution 
of K~XVIII and Cl~XVII lines\footnote{The fact that emission near 3.5~keV from Galactic Center region
is consistent with adding K~XVIII lines at 3.47 and 3.51~keV was first mentioned in~\cite{Riemer-Sorensen:14}.
\cite{Boyarsky:14a} also mentioned that fact and showed that using Galactic Center data alone it was not 
possible to neither claim the existence an unidentified spectral line on top of the element lines, 
nor constrain it.} and b) extra line from M31 center seen by~\cite{Boyarsky:14a} can be lowered 
to $< $90\% confidence by adjusting X-ray continuum over small energy range near the line (3-4~keV).

The criticism of~\cite{Jeltema:14} have stimulated the immediate comment of~\cite{Boyarsky:14c}.
Here, claim b) of~\cite{Jeltema:14} is repudiated by showing both that X-ray continuum of~\cite{Jeltema:14}
selected at 3-4~keV is significantly overestimated at larger energies, and that the extra line flux is at least 
an order of magnitude less than expected from astrophysical lines near 3.5~keV. 
The other concern a) of~\cite{Jeltema:14} has recently been commented in~\cite{Bulbul:14b} showing that the 
analysis of~\cite{Jeltema:14} suffers from their use of the approximate atomic data from ATOMDB~\cite{AtomDB} 
website. In contrast, full version of ATOMDB used by~\cite{Bulbul:14a,Boyarsky:14b,Boyarsky:14c} leads to 
significant lack of astrophysical emission to explain observed line at $\sim 3.5$~keV in 
combined dataset of galaxy clusters of~\cite{Bulbul:14a}. 
These comments, in turn, have been replied in~\cite{Jeltema:14b}. By using larger energy range proposed 
by~\cite{Boyarsky:14a,Boyarsky:14b},~\cite{Jeltema:14b} recovered the initial result of~\cite{Boyarsky:14a}
about the 3.5~keV line significance; however, unlike~\cite{Boyarsky:14a},~\cite{Jeltema:14b} obtained much 
more significant line-like negative residual below 3~keV. The origin of this discrepancy is to be found. 
A probable reason is the slope of instrumental component seen in Fig.~2 of~\cite{Jeltema:14b} from
its fiducial value $E^{-0.2}$, see e.g.~\cite{Lumb:02}, which can, due to $\sim 5\%$ dip in effective area 
(see e.g. Fig.~1 of~\cite{Jeltema:14b}), mimic such a large negative residual. On the other 
hand, by using full ATOMDB version~\cite{Jeltema:14b} presented an additional argument of the initial 
claim of~\cite{Jeltema:14} based on new Ca~XIX/Ca~XX line ratios, so more detailed investigation (including 
e.g. systematic uncertainties on ion emissivities) is required to finally resolve this issue.

An alternative approach is to study the \emph{line morphology}. In~\cite{Carlson:14}, \xmm\ 
observations of the central part of the Perseus cluster and the Galactic Center have been analyzed. \cite{Carlson:14}
collected \emph{all} events (either cosmic or instrumental origin) in narrow energy ranges 
(roughly corresponding to the energy resolution), and looked for the best-fit approximation with 
the rescaled continuum obtained from several adjacent line-free bands. The main result of~\cite{Carlson:14}
is that adding decaying dark matter distribution from a \emph{smooth} DM profile 
(Navarro-Frenk-White, Einasto, Burkert) does not improve the fit quality in both objects. 
In addition,~\cite{Carlson:14} demonstrated that distribution of the events in 
3.45-3.6~keV bands correlates with that in the energy bands of strong astrophysical emission, rather than 
with that in line-free energy bands. Based on these findings, \cite{Carlson:14} claimed the exclusion of decaying 
dark matter origin of 3.5~keV in the Galactic Center and the Perseus cluster. 

2. Linear scaling of line position with the redshift observed by~\cite{Bulbul:14a,Boyarsky:14a} is an 
important argument \emph{against} the instrumental origin of $\sim 3.5$~keV line. The fact that this line
has not been found in long blank-sky dataset of~\cite{Boyarsky:14a} provides additional evidence against
its instrumental origin.

3. The authors of~\cite{Bulbul:14a,Boyarsky:14a,Boyarsky:14b} argue that all basic properties of detected 
$\sim 3.5$~keV line -- its position, line strength, scaling with redshift, angular distribution inside 
extended objects (Perseus cluster outskirts, center vs off-center of Andromeda galaxy, 
Galactic Center vs blank-sky dataset), scaling among different objects, and even its non-observation in some 
datasets of~\cite{Bulbul:14a,Boyarsky:14a} -- are all consistent with an explanation
in terms of \emph{radiatively decaying dark matter line}. 

The predictive power of the decaying dark matter 
scenario motivated several groups of researchers~\cite{Riemer-Sorensen:14,Jeltema:14,
Malyshev:14,Anderson:14,Urban:14,Carlson:14} 
to study X-ray spectra of different dark matter-dominated 
objects. Their studies are summarized in Table~\ref{tab:line-properties}. At the moment, no further confirmation
of decaying dark matter origin after the papers~\cite{Bulbul:14a,Boyarsky:14a,Boyarsky:14b} has been 
presented. While~\cite{Riemer-Sorensen:14,Jeltema:14,Urban:14,Carlson:14} found a line at 
$\sim 3.5$~keV (though interpreted it as a sum of astrophysical lines), \cite{Malyshev:14}
and \cite{Anderson:14} have not detected the line in combined datasets of dwarf spheroidal galaxies (dSphs) 
and spiral galaxies, respectively, placing only upper bounds on dark matter lifetime. Non-detection
of the line at 3.55~keV by \cite{Malyshev:14} is still consistent with decaying dark matter hypothesis;
to rule it out, even quoting \cite{Malyshev:14}\footnote{Given large uncertainties in dark matter modeling,
the obtained bounds are usually (see e.g.~\cite{Boyarsky:12c,Boyarsky:07a}) diluted by an extra factor of 2, 
contrary to~\cite{Malyshev:14}.}, one needs to increase the sensitivity by a factor 
$\sim 2$ (which means a factor $\sim 4$ increase of exposure assuming similar dark matter column density).
On the other hand, non-observation of $\sim 3.5$~keV line in the dataset of~\cite{Anderson:14} 
(see their Fig.~4) may be interpreted as a tension with decaying dark matter hypothesis and therefore motivates 
more detailed study. According to~\cite{Iakubovskyi:13}, where combined dataset of galaxies with comparable 
exposure has been analyzed, at exposures larger than $\sim 10$~Msec \emph{line-like} systematic errors start to 
 dominate over statistical errors. As a result, usual method of determination of continuum level (by simply
 minimizing $\chi^2$, as~\cite{Anderson:14} did) is no longer appropriate. Previous studies of line in M31 
 center~\cite{Boyarsky:14a,Jeltema:14,Boyarsky:14c} shows that precise determination of continuum level is 
 important to quantify the intensity of 3.5~keV line. Therefore, the only way to put \emph{robust} exclusions
 to line intensity is to perform continuum modeling in a way similar to~\cite{Iakubovskyi:13}: to decrease 
 level of continuum \emph{below} the best-fit to ensure absence of significant negative residuals, 
 and to add systematic errors to account non-Gaussian distribution of positive residuals. According to 
 Fig.~5.26 of~\cite{Iakubovskyi:13}, such analysis would produce 3$\sigma$ upper bounds for 3.5~keV line flux
 close to $\sim 1.5\times 10^{-6}$ ph/sec/cm$^{2}$ per \xmm\ field-of-view, still consistent with line 
 observation in M31 center~\cite{Boyarsky:14a}.

Although the results of~\cite{Bulbul:14a,Boyarsky:14a} are formulated for a specific dark matter candidate -- 
right-handed (sterile) neutrinos (see~\cite{Boyarsky:09a,Boyarsky:12c} for recent reviews), they can be 
applied for \emph{any}
type of radiatively decaying dark matter, see 
e.g.~\cite{Ishida:14a,Higaki:14,Jaeckel:14,Czerny:14,Lee:14a,Abazajian:14,Krall:14,
ElAisati:14,Kong:14,Frandsen:14,Baek:14a,Nakayama:14a,Choi:14,Shuve:14,Kolda:14,Allahverdi:14,
Dias:14,Bomark:14,PeiLiew:14,Nakayama:14b,Kang:14,Demidov:14,Queiroz:14,Babu:14,Prasad:14,Rosner:14,
Lee:14b,Robinson:14,Baek:14b,Nakayama:14c,Chakraborty:14,Chen:14,Ishida:14b,Geng:14,Abada:14a,
Chiang:14,Dutta:14,Okada:14c,Haba:14,Henning:14,Farzan:14,Faisel:14}.
The difference among these models can be further probed by:
\begin{itemize}
 \item changes in line morphology due to non-negligible initial dark matter velocities, see 
 e.g.~\cite{Maccio:12b,Lovell:13a};
 \item other astrophysical tests such as Ly-$\alpha$ method, see e.g.~\cite{Viel:13,Abazajian:14,Merle:14};
 \item search of ``smoking gun'' signatures in accelerator experiments, see e.g.~\cite{Bonivento:13} for 
 the minimal neutrino extension of the Standard Model, $\nu$MSM~\cite{Asaka:05a,Asaka:05b,Boyarsky:09a}.
\end{itemize}
 
4. Recently proposed alternatives to radiatively decaying dark matter currently include decay of excited dark 
matter states~\cite{Finkbeiner:14,Cline:14,Okada:14b,Cline:14b,Boddy:14,Schutz:14,Cline:14c}, 
annihilating dark matter~\cite{Dudas:14,Frandsen:14,Baek:14b}, 
dark matter decaying into axion-like particles with further conversion to photons in magnetic 
field~\cite{Cicoli:14,Conlon:14a,Conlon:14b,Alvarez:14}. These models predict \emph{substantial} difference in 
$\sim 3.5$~keV line morphology compared to the radiatively decaying dark matter. For example, their line profiles
should be more concentrated towards the centers of dark matter-dominated objects due to larger dark matter 
density (for exciting and annihilating dark matter) and larger magnetic fields (for magnetic field conversion
of axion-like particles). Further non-observation of the $\sim 3.5$~keV line in outskirts of dark matter-dominated
objects would therefore an agrument in favour of these models.
 
\section*{\sc Conclusions and future directions}

\indent \indent New emission line at $\sim$3.5~keV in spectra of galaxy clusters and central parts of Andromeda 
galaxy recently reported by~\cite{Bulbul:14a,Boyarsky:14a}
remains unexplained in terms of astrophysical emission lines or instrumental features (see however recent 
works~\cite{Carlson:14,Jeltema:14b}). 
Its properties are consistent with radiatively decaying dark matter and other intereresting scenarios
(such as, exciting dark matter, annihilating dark matter and dark matter decaying into axion-like particles
further converted in cosmic magnetic fields) motivated by various particle physics 
extensions of the Standard Model. In case of radiatively decaying dark matter, further detection of new 
emission line in other objects would lead to direct detection of new physics. Specially dedicated observations 
by existing X-ray missions (such as \xmm, \chan, \suza) still allow such detection (see e.g.~\cite{Lovell:14}) 
although one should take detailed care on various systematic effects that could mimic or hide the new line. 

The alternative is to use new better instruments. The basic requirements for such instruments -- higher 
\emph{grasp} (the product of field-of-view and effective area) and better \emph{spectral resolution}
\footnote{\emph{Grating} spectrometers such as \chan/HETGS have excellent spectral resolution for \emph{point}
sources; however, for extended ($\gtrsim$1~arcmin) sources their spectral resolution usually degrades to that 
for existing imaging spectrometers, see e.g.~\cite{Dewey:02}.} -- 
have first formulated in~\cite{Boyarsky:06f}. The imaging spectrometer on-board new X-ray mission 
\textit{Astro-H}~\cite{Takahashi:12} scheduled to launch in 2015 meets only second requirement 
having energy resolution by an order of magnitude better ($\sim 5$~eV) compared to existing instruments. 
This will allow \textit{Astro-H}
to precisely determine the line position in brightest objects with prolonged observations
(according to~\cite{Bulbul:14a}, a 1~Msec observation of the Perseus cluster is required) and thus 
\textit{Astro-H} will finally close the question whether the new line is from new physics or from 
(anomalously enhanced) astrophysical emission. Other interesting possibility is proposed in~\cite{Boyarsky:14a}: 
one should see decaying dark matter signal from the Milky Way halo in \emph{every} \textit{Astro-H} observation, 
therefore their compination could also reveal the radiatively decaying dark matter nature of new line.
Another possibility is to use planned \textit{LOFT} mission~\cite{Zane:14} which high grasp and moderate energy 
resolution would allow to detect new line at much smaller intensities~\cite{Neronov:13}. 
Finally, an ``ultimate'' imaging spectrometer proposed in e.g.~\cite{Boyarsky:12c}
would reveal the detailed structure of $\sim$3.5~keV line.

\section*{\sc acknowledgement}
\indent \indent This work was supported by part by the Swiss National Science Foundation grant 
SCOPE IZ7370-152581, the Program of Cosmic Research of the National Academy of 
Sciences of Ukraine, the State Programme of Implementation of Grid Technology in Ukraine 
and the grant of President of Ukraine for young scientists.

%


\let\jnlstyle=\rm\def\jref#1{{\jnlstyle#1}}\def\aj{\jref{AJ}}
  \def\araa{\jref{ARA\&A}} \def\apj{\jref{ApJ}\ } \def\apjl{\jref{ApJ}\ }
  \def\apjs{\jref{ApJS}} \def\ao{\jref{Appl.~Opt.}} \def\apss{\jref{Ap\&SS}}
  \def\aap{\jref{A\&A}} \def\aapr{\jref{A\&A~Rev.}} \def\aaps{\jref{A\&AS}}
  \def\azh{\jref{AZh}} \def\baas{\jref{BAAS}} \def\jrasc{\jref{JRASC}}
  \def\memras{\jref{MmRAS}} \def\mnras{\jref{MNRAS}\ }
  \def\pra{\jref{Phys.~Rev.~A}\ } \def\prb{\jref{Phys.~Rev.~B}\ }
  \def\prc{\jref{Phys.~Rev.~C}\ } \def\prd{\jref{Phys.~Rev.~D}\ }
  \def\pre{\jref{Phys.~Rev.~E}} \def\prl{\jref{Phys.~Rev.~Lett.}}
  \def\pasp{\jref{PASP}} \def\pasj{\jref{PASJ}} \def\qjras{\jref{QJRAS}}
  \def\skytel{\jref{S\&T}} \def\solphys{\jref{Sol.~Phys.}}
  \def\sovast{\jref{Soviet~Ast.}} \def\ssr{\jref{Space~Sci.~Rev.}}
  \def\zap{\jref{ZAp}} \def\nat{\jref{Nature}\ } \def\iaucirc{\jref{IAU~Circ.}}
  \def\aplett{\jref{Astrophys.~Lett.}}
  \def\apspr{\jref{Astrophys.~Space~Phys.~Res.}}
  \def\bain{\jref{Bull.~Astron.~Inst.~Netherlands}}
  \def\fcp{\jref{Fund.~Cosmic~Phys.}} \def\gca{\jref{Geochim.~Cosmochim.~Acta}}
  \def\grl{\jref{Geophys.~Res.~Lett.}} \def\jcp{\jref{J.~Chem.~Phys.}}
  \def\jgr{\jref{J.~Geophys.~Res.}}
  \def\jqsrt{\jref{J.~Quant.~Spec.~Radiat.~Transf.}}
  \def\memsai{\jref{Mem.~Soc.~Astron.~Italiana}}
  \def\nphysa{\jref{Nucl.~Phys.~A}} \def\physrep{\jref{Phys.~Rep.}}
  \def\physscr{\jref{Phys.~Scr}} \def\planss{\jref{Planet.~Space~Sci.}}
  \def\procspie{\jref{Proc.~SPIE}} \let\astap=\aap \let\apjlett=\apjl
  \let\apjsupp=\apjs \let\applopt=\ao \def\jcap{\jref{JCAP}}

\end{document}